# A conservation rule for constructing bibliometric network matrices

Gangan Prathap[1] and Somenath Mukherjee[2]

[1]*A P J Abdul Kalam Technological University, Thiruvananthapuram, Kerala, India 695016. e-mail: gangan_prathap@hotmail.com*

[2]*CSIR Central Mechanical Engineering Research Institute, Durgapur, Bengal, India 713209. e-mail: somenath@cmeri.res.in*

**Abstract**

The social network analysis of bibliometric data needs matrices to be recast in a network framework. In this paper we argue that a simple conservation rule requires that this should be done only using fractional counting so that conservation at the paper level will be faithfully reproduced at higher levels of aggregation (i.e. author, institute, country, journal etc.) of the complex network.

## Introduction

Quite early in the development of bibliometrics as a field of enquiry the network properties were anticipated (De Solla Price, 1965; Kessler, 1963; Small, 1973). There is now a pronounced trend towards developing more sophisticated indicators for scholarly performance evaluation (Bollen et al. 2006; Leydesdorff 2009) using social-network analysis (Pinski and Narin 1976; Brin and Page 1998; Bergstrom 2007). The social network analysis of bibliometric data needs matrices to be recast in a network framework. (e.g., Börner, Chen, & Boyack, 2003; Milojević, 2014; Van Eck & Waltman, 2014; Zhao & Strotmann, 2015).

A key issue in constructing a bibliometric network is whether a full counting or a fractional counting approach is to be used (Batagelj & Cerinšek, 2013; Park, Yoon, & Leydesdorff, 2016). Perianes-Rodriguez, Waltman, & van Eck (2016) argue that the fractional counting method is preferable over the full counting method. In this paper we further argue that a simple conservation rule requires that only fractional counting faithfully reproduces the conservation rule introduced at the paper level at higher levels of aggregation (i.e. author, institute, country, journal etc.) of the complex network.

**The conservation rule at the paper and network levels**

Table 1 shows an instance where four authors (a1, a2, a3 and a4) publish three papers (p1, p2 and p3), an example taken from Leydesdorff & Park (2016). The top half of the table shows how the credit for the papers is assigned to the individual authors under the full counting scheme and the bottom half shows this for the fractional counting scheme. It is clearly seen that under full counting, the conservation law is violated leading to an inflation of paper count from 3 to 7. However, in fractional counting, conservation of the total is also maintained at the network level as we shall demonstrate below. Note that in fractional counting, any rule (here an equal credit to all authors rule is shown) can be used as long as the conservation rule is followed for each paper (that is each column must add up to 1). Let the authorship matrix at the paper level be designated by $\mathbf{A}$ where the elements are $a_{ij}$, for the contribution of author $i$ to paper $j$, there being a total of $I$ authors and $J$ papers.

Let us represent the authorship matrix at the network level by $\mathbf{B}$ where $\mathbf{B} = \mathbf{A}\mathbf{A}^T$. Table 2 shows how the authorship matrix at the network level is constructed when full counting or fractional counting is used at the paper level. We see immediately that only fractional counting is able to conserve the total number of papers. Full counting has now inflated the count at the network level to 17.

Another lesson that emerges from Table 2 is that the diagonal terms have to be non-zero for conservation to be true. A simple thought experiment will establish this. Consider the case of $I$ authors all of whom have been single authors of a paper each, *i.e.* author $i$ is the single author of paper *i*. The **A** matrix is a diagonal matrix and so would the **B** matrix be; that is, all diagonal terms will have to be conserved.

In the next section a simple algebraic procedure shows that whatever the number of authors *I*, after the $\mathbf{AA}^T$ operation at the network level, the total number of papers will remain at *J*.

**A simple proof of the conservation rule at the paper and network levels**

>  **A** is an $I \times J$ matrix, $\mathbf{A} = [a_{ij}]_{I \times J}$ wherein the *I* elements of each of its *J* columns add up to unity, *i.e.* $\sum_{i=1}^{I} a_{ij} = 1; \quad j = 1, 2 \ldots J$.
>
> A new matrix **B** is given as $\mathbf{B} = \mathbf{AA}^T$. Show that $\sum_{i=1}^{I}\sum_{j=1}^{I} b_{ij} = J$.
>
> *Proof*
> $$\mathbf{A}^T = [a'_{ij}]_{J \times I} \quad \text{where } a'_{ij} = a_{ji} \quad \ldots \ldots \ldots \ldots (1)$$
> Thus $\mathbf{B} = \mathbf{AA}^T = [a_{ij}]_{I \times J} \cdot [a'_{ij}]_{J \times I} = [b_{ij}]_{I \times I}$
> wherein
> $$b_{ij} = \sum_{k=1}^{J} a_{ik} \cdot a'_{kj} = \sum_{k=1}^{J} a_{ik} \cdot a_{jk} \quad (\because \text{condition (1))}$$
> Thus
> $$\sum_{i=1}^{I}\sum_{j=1}^{I} b_{ij} = \sum_{i=1}^{I} b_{ii} + \sum_{i=1}^{I}\sum_{\substack{j=1 \\ i \neq j}}^{I} b_{ij} = \sum_{i=1}^{I}\sum_{k=1}^{J} (a_{ik})^2 + \sum_{i=1}^{I}\sum_{\substack{j=1 \\ i \neq j}}^{I} \sum_{k=1}^{J} a_{ik} \cdot a_{jk}$$
> $$= \sum_{k=1}^{J} \left[ \sum_{i=1}^{I} a_{ik}^2 + \sum_{i=1}^{I}\sum_{\substack{j=1 \\ i \neq j}}^{I} a_{ik} \cdot a_{jk} \right] = \sum_{k=1}^{J} \left[ \sum_{p=1}^{I} a_{pk} \right]^2 \quad \ldots \ldots (2)$$
>
> Since by the given condition, $\sum_{p=1}^{I} a_{pk} = 1; \quad k = 1, 2 \ldots J$ one gets from (2) above,
> $$\sum_{i=1}^{I}\sum_{j=1}^{I} b_{ij} = \sum_{k=1}^{J} 1^2 = 1 + 1 + 1 \ldots \text{upto } J \text{ times} = J \quad \ldots \ldots \ldots (3)$$
>
> Note that in statement (2), the orders of indices, *ij* and *ji* (two possible *permutations* for $i \neq j$) are both accommodated. Since the matrix **B** is *symmetric* ($b_{ij} = b_{ji}$), only one *combination* of indices *i* and *j* (without change of order) may be permitted (*i.e. either* with $i < j$ *or* with $i > j$), with a factor 2 incorporated in the appropriate expressions,

$$\sum_{i=1}^{I}\sum_{j=1}^{I}b_{ij} = \sum_{i=1}^{I}b_{ii} + 2\times\sum_{i=1}^{I}\sum_{\substack{j=1\\i<j}}^{I}b_{ij} = \sum_{i=1}^{I}\sum_{k=1}^{J}(a_{ik})^2 + 2\times\sum_{i=1}^{I}\sum_{\substack{j=1\\i<j}}^{I}\sum_{k=1}^{J}a_{ik}.a_{jk}$$

$$= \sum_{k=1}^{J}\left[\sum_{i=1}^{I}a_{ik}^{\,2} + 2\times\sum_{i=1}^{I}\sum_{\substack{j=1\\i<j}}^{I}a_{ik}.a_{jk}\right] = \sum_{k=1}^{J}\left[\sum_{p=1}^{I}a_{pk}\right]^2 = \sum_{k=1}^{J}1^2 = J \quad \ldots\ldots(4)$$

**Concluding remarks**

When graph theoretic procedures from social network analysis are applied to bibliometric data one must take care to maintain the conservation principle. In this paper we have shown from an empirical example as well as a formal proof that the conservation rule requires that this should be done only using fractional counting so that conservation at the paper level will be faithfully reproduced at higher levels of aggregation (*i.e.* author, institute, country, journal etc.) of the complex network.

**References**

Batagelj, V. & Cerinšek, M. (2013). On bibliographic networks. *Scientometrics*, 96(3), 845–864.

Bergstrom, C. (2007). Eigenfactor: Measuring the value and prestige of scholarly journals. *College & Research Libraries News*, 68, 314.

Bollen, J., Rodriguez, M. A., & Van de Sompel, H. (2006). Journal status. *Scientometrics*, 69(3), 669–687.

Börner, K., Chen, C., & Boyack, K.W. (2003). Visualizing knowledge domains. Annual Review of Information Science and Technology, 37(1), 179–255.

Brin, S., & Page, L. (1998). The anatomy of a large-scale hypertextual Web search engine. *Computer Networks and ISDN Systems*, 30(1-7), 107-117.

De Solla Price, D.J. (1965). Networks of scientific papers. *Science*, 149, 510–515.

Kessler, M.M. (1963). Bibliographic coupling between scientific papers. *American Documentation*, 14(1), 10–25.

Leydesdorff, L. (2009). How are New Citation-Based Journal Indicators Adding to the Bibliometric Toolbox? *Journal of the American Society for Information Science and Technology*, 60(7), 1327–1336.

Leydesdorff, L., & Park, H. W. (2016). Full and Fractional Counting in Bibliometric Networks. (https://arxiv.org/ftp/arxiv/papers/1611/1611.06943.pdf )

Milojević, S. (2014). Network analysis and indicators. In Y. Ding, R. Rousseau, & D. Wolfram (Eds.), Measuring scholarly impact: Methods and practice (pp. 57–82). Springer.

Park, H.W., Yoon, J., & Leydesdorff, L. (2016). The normalization of co-authorship networks in the bibliometric evaluation: The government stimulation programs of China and Korea. *Scientometrics*, 109(2), 1017–1036.

Perianes-Rodriguez, A., Waltman, L., & van Eck, N. J. (2016). Constructing bibliometric networks: A comparison between full and fractional counting. *Journal of Informetrics*, 10(4), 1178-1195.

Pinski, G., & Narin, F. (1976). Citation influence for journal aggregates of scientific publications: Theory, with application to the literature of physics. *Information Processing and Management*, 12(5), 297–312.

Small, H. (1973). Co-citation in the scientific literature: A new measure of the relationship between two documents. Journal of the American Society for Information Science, 24(4), 265–269.

Van Eck, N.J., & Waltman, L. (2014). Visualizing bibliometric networks. In Y. Ding, R. Rousseau, & D. Wolfram (Eds.), Measuring scholarly impact: Methods and practice (pp. 285–320). Springer.

Zhao, D., & Strotmann, A. (2015). Analysis and visualization of citation networks. Morgan & Claypool Publishers.

Table 1. The authorship matrix when Full counting or Fractional counting is used at the paper level.

| A matrix | Full counting | | | |
|---|---|---|---|---|
| | **p1** | **p2** | **p3** | **Total** |
| **a1** | 1 | 1 | 0 | 2 |
| **a2** | 1 | 0 | 1 | 2 |
| **a3** | 1 | 1 | 0 | 2 |
| **a4** | 0 | 0 | 1 | 1 |
| **Total** | 3 | 2 | 2 | **7** |
| A matrix | Fractional counting | | | |
| | **p1** | **p2** | **p3** | **Total** |
| **a1** | 0.333 | 0.500 | 0.000 | 0.833 |
| **a2** | 0.333 | 0.000 | 0.500 | 0.833 |
| **a3** | 0.333 | 0.500 | 0.000 | 0.833 |
| **a4** | 0.000 | 0.000 | 0.500 | 0.500 |
| **Total** | 1 | 1 | 1 | **3** |

Table 2. The authorship matrices at the paper and network levels when Full counting or Fractional counting is used at the paper level.

| | | | | $A^T$ matrix | a1 | a2 | a3 | a4 | Total |
|---|---|---|---|---|---|---|---|---|---|
| | **Full counting** | | | p1 | 1 | 1 | 1 | 0 | 3 |
| | | | | p2 | 1 | 0 | 1 | 0 | 2 |
| | | | | p3 | 0 | 1 | 0 | 1 | 2 |
| A matrix | p1 | p2 | p3 | B matrix | a1 | a2 | a3 | a4 | Total |
| a1 | 1 | 1 | 0 | a1 | 2 | 1 | 2 | 0 | 5 |
| a2 | 1 | 0 | 1 | a2 | 1 | 2 | 1 | 1 | 5 |
| a3 | 1 | 1 | 0 | a3 | 2 | 1 | 2 | 0 | 5 |
| a4 | 0 | 0 | 1 | a4 | 0 | 1 | 0 | 1 | 2 |
| Total | 3 | 2 | 2 | Total | 5 | 5 | 5 | 2 | **17** |
| | | | | $A^T$ matrix | a1 | a2 | a3 | a4 | Total |
| | **Fractional counting** | | | p1 | 0.333 | 0.333 | 0.333 | 0.000 | 1 |
| | | | | p2 | 0.500 | 0.000 | 0.500 | 0.000 | 1 |
| | | | | p3 | 0.000 | 0.500 | 0.000 | 0.500 | 1 |
| A matrix | p1 | p2 | p3 | B matrix | a1 | a2 | a3 | a4 | Total |
| a1 | 0.333 | 0.500 | 0.000 | a1 | 0.361 | 0.111 | 0.361 | 0.000 | 0.833 |
| a2 | 0.333 | 0.000 | 0.500 | a2 | 0.111 | 0.361 | 0.111 | 0.250 | 0.833 |
| a3 | 0.333 | 0.500 | 0.000 | a3 | 0.361 | 0.111 | 0.361 | 0.000 | 0.833 |
| a4 | 0.000 | 0.000 | 0.500 | a4 | 0.000 | 0.250 | 0.000 | 0.250 | 0.500 |
| Total | 1 | 1 | 1 | Total | 0.833 | 0.833 | 0.833 | 0.500 | **3** |